\documentclass[twocolumn,preprintnumbers,amsmath,nofootinbib,amssymb]{revtex4}
\usepackage{amssymb}
\usepackage{latexsym}
\usepackage{color}
\usepackage{graphicx}
\usepackage{hyperref}
\usepackage{color}

\def\nn{\nonumber\\}
\newcommand{\f}[2]{\frac{#1}{#2}}
\def\be{\begin{equation}}
\def\ee{\end{equation}}
\def\bea{\begin{eqnarray}}
\def\eea{\end{eqnarray}}
\begin{document}
\title{Gravitational Collapse in Rastall Gravity}
\author{A. H. Ziaie\footnote{ah.ziaie@riaam.ac.ir}, H. Moradpour\footnote{h.moradpour@riaam.ac.ir}, S. Ghaffari\footnote{sh.ghaffari@riaam.ac.ir}}
\address{$^1$ Research Institute for Astronomy and Astrophysics of Maragha (RIAAM), P.O. Box
55134-441, Maragha, Iran}
\begin{abstract}
We study spherically symmetric gravitational collapse of a homogeneous perfect fluid in Rastall gravity. Considering a linear equation of state (EoS) for the fluid profiles, we examine the conditions under which the collapse scenario could end in a spacetime singularity. Depending on the model parameters, the singularity could be either naked or covered by a horizon. We find that a non-vanishing Rastall parameter could affect the formation of apparent horizon so that, naked singularities may form for those values of EoS parameter for which a homogeneous perfect fluid collapse in general relativity (GR) terminates at the black hole formation. Hence the visibility of the resulting singularity depends on the Rastall parameter. The solutions we obtain respect the weak energy condition (WEC) which is crucial for physical validity of the model.
\end{abstract}

\maketitle
\section{Introduction}
The question of final fate of a continual gravitational collapse is of great significance in gravitation theory and astrophysics
today. When a star with mass many times than the solar mass exhausts its internal nuclear fuel, it is believed that it undergoes an endless gravitational collapse without having any final equilibrium state such as a neutron star or a white dwarf. In the context of GR, the Hawking and Penrose singularity theorems predict that under physically reasonable circumstances, the star goes on shrinking in its radius, reaching higher and higher densities and finally the collapse process culminates in the formation a spacetime singularity, where densities and spacetime curvatures become infinitely large and blow up~\cite{HAWPENST}. However, these theorems do not provide much information on the nature of singularities they predict.
\par
Since the last decades, the subject of gravitational collapse has attracted much attention and the studies having been made in this area show that the resulting singularity could be covered by a spacetime event horizon (black hole) or may be observed by the external universe (naked singularity)~\cite{ColReview}. The latter could in principle expose us super dense regimes of extreme gravity with ultra small length scales, providing then a laboratory for detecting quantum gravity effects near these regimes. While, in the former, such regimes are causally disconnected from observers on the outside of the universe, according to the Cosmic Censorship Conjecture (CCC)~\cite{CCCREF}. Although there is no conclusive proof or disproof of CCC up until now, several attempts have been made to find a number of counterexamples which showed there are naked singularities occurring at the center of spherically symmetric dust, perfect fluids or radiation shells~\cite{Joshibook}. Work along this line has been also extended to modified gravity theories, where its shown that naked singularities could form depending on different aspects of the theory, see e.g.,~\cite{ALTGRNAKED}. It is therefore of interest to study other gravitational theories whose geometrical attributes (not present in GR) may affect the final fate of the collapse scenario. One way to modify GR is to consider a non-minimal coupling between geometry and matter fields. Due to this assumption, the ordinary energy-momentum conservation law ($\nabla_\mu T^\mu_{\,\,\,\, \nu}=0$) is not valid anymore \cite{od1,od2,cmc,cmc1,cmc2,rastall,genras}. The origin of this idea is rooted in the Rastall's work \cite{rastall} which is in good agreement with various observational data and theoretical expectations \cite{dmy}. In the present work, we are motivated to consider a simple model for gravitational collapse of a homogeneous perfect fluid in Rastall gravity and study under what conditions on model parameters, a naked singularity may form. 
\section{Field equations of Rastall gravity}
According to the original idea of Rastall~\cite{rastall}, the vanishing of covariant divergence of the matter energy-momentum tensor is no longer valid and this vector field is proportional to the covariant derivative of the Ricci curvature scalar as
\be\label{CoVDiv}
\nabla_{\mu}T^{\mu}_{\,\,\,\nu}=\lambda\nabla_{\nu}{\mathcal R},
\ee
where $\lambda$ is the Rastall parameter. The Rastall field equations are then given by~\cite{rastall,FESRASPLB}
\be\label{RastallFES}
G_{\mu\nu} +\gamma g_{\mu\nu}{\mathcal R}=\kappa T_{\mu\nu},
\ee
where $\gamma=\kappa\lambda$ is the Rastall dimensionless parameter and $\kappa$ being the Rastall gravitational coupling constant. The above equation can be rewritten in an equivalent form as
\be\label{FESEquiv}
G_{\mu\nu}=\kappa S_{\mu\nu},~~~~~S_{\mu\nu}=T_{\mu\nu}-\f{\gamma T}{4\gamma-1}g_{\mu\nu},
\ee
where $S_{\mu\nu}$ is the effective energy momentum tensor whose components are given by~\cite{Effemtras}
\bea
S^0_{\,\,0}\equiv-\rho^{\rm eff}=-\f{(3\gamma-1)\rho+\gamma(p_r+2p_t)}{4\gamma-1},\label{EFFEMT1}\\
S^1_{\,\,1}\equiv p_r^{\rm eff}=\f{(3\gamma-1)p_r+\gamma(\rho-2p_t)}{4\gamma-1},\label{EFFEMT2}\\
S^2_{\,\,2}=S^3_{\,\,3}\equiv p_t^{\rm eff}=\f{(2\gamma-1)p_t+\gamma(\rho-p_r)}{4\gamma-1}.\label{EFFEMT3}
\eea
It is noteworthy that in the limit of $\lambda\rightarrow0$ the standard GR is recovered. Moreover for an electromagnetic field source we get $S_{\mu\nu}=T_{\mu\nu}$ leading to $G_{\mu\nu}=\kappa T_{\mu\nu}$. Therefore, the GR solutions for $T=0$, or equivalently $R=0$, are also valid the Rastall gravity~\cite{validity1,validity2}.
\section{Solutions to the field equations}
\begin{figure}
	\includegraphics[scale=0.32]{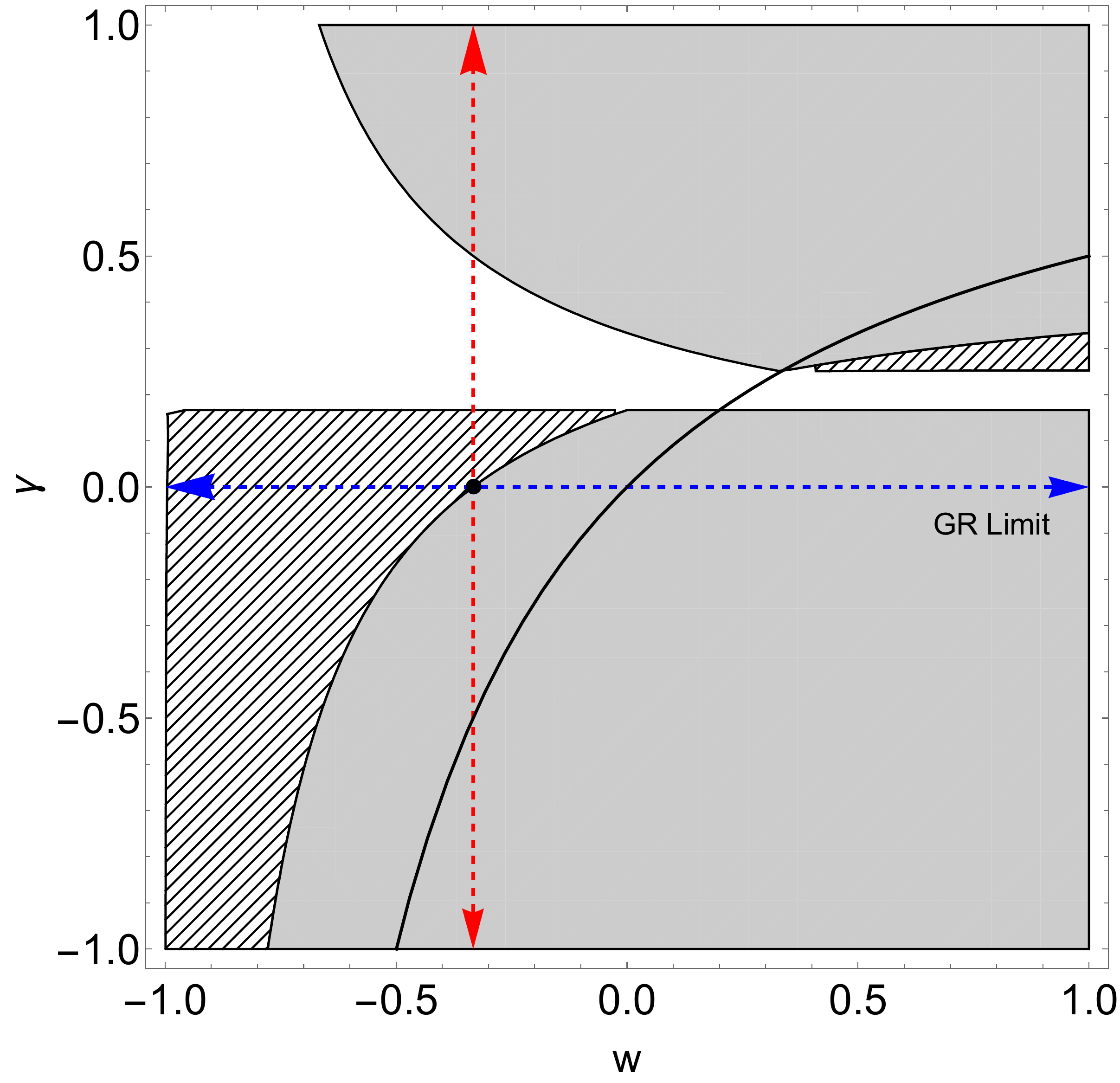}
	\caption{The allowed values for EoS and Rastall parameters that determine the formation (gray region) or failure (shaded region) of apparent horizon. The homogeneous collapse scenario for any pair of $(\gamma,w)$ parameters chosen from gray (shaded) region leads to black hole (naked singularity) formation. Red and blue dashed arrows correspond to $w=-\f{1}{3}$ and $\gamma=0$, respectively. Any point $(\gamma,w)$ lying of the solid curve satisfies equation
		$w-\gamma(1+w)=0$. The white region is not allowed for the model parameters.}\label{FIGWG1}
\end{figure}

In the present section, we seek for a class of homogeneous collapse solutions which lead to the formation of a spacetime singularity. However, if we assume fewer symmetries namely considering inhomogeneities or anisotropies within the collapsing object, there would be a lack of physically reasonable exact solutions available due to the intrinsic mathematical complexities. We therefore restrict our analysis to a homogeneous and isotropic interior line element which represents a spatially flat
FLRW geometry
\be\label{FLRW}
ds^2=-dt^2+a^2(t)dr^2+R^2(t,r)d\Omega^2,
\ee
where $R(t,r)=ra(t)$ is the physical radius of the collapsing body, with $a(t)$ being the scale factor and $d\Omega^2$ is the standard line element
on the unit 2-sphere. The generalized Friedman-type field equations for an isotropic source ($T^{\mu}_{\,\,\nu}={\rm diag}(-\rho,p,p,p)$) then read
\bea
(3-12\gamma)H^2-6\gamma\dot{H}=\kappa\rho,\label{RasFES}\\
(12\gamma-3)H^2+(6\gamma-2)\dot{H}=\kappa p,\label{RasFES1}
\eea
where $\kappa=2(4\gamma-1)\kappa_G/(6\gamma-1)$, $H=\f{\dot{a}}{a}$ is the collapse rate and $\kappa_G=4\pi G$. Applying the Bianchi identity on Eq. \ref{FESEquiv} leaves us with the following continuity equation in Rastall gravity, as
\be\label{conteq}
\left(\f{3\gamma-1}{4\gamma-1}\right)\dot{\rho}+\left(\f{3\gamma}{4\gamma-1}\right)\dot{p}+3H(\rho+p)=0.
\ee
Next, we proceed to find an exact solution assuming a linear EoS $p=w\rho$, where $w$ is the barotropic index. Therefore, from  Eqs. (\ref{RasFES}) and (\ref{RasFES1}) we get the following differential equation for the collapse rate as (we set the units so that $2\kappa_G=1$)
\be\label{diffeq}
3(1+w)(4\gamma-1)H^2+2(3\gamma(1+w)-1)\dot{H}=0.
\ee
Integrating twice gives the sale factor as
\be\label{scfac}
a(t)=C_2\left[3t(1+w)(4\gamma-1)-2C_1(3\gamma(1+w)-1)\right]^{n},
\ee
where
\be\label{nexponent}
n=\f{-2+6\gamma(1+w)}{3(1+w)(4\gamma-1)},
\ee
and $C_1$ and $C_2$ are the integration constants. These two constants can be straightforwardly determined by setting $a(t_0)=a_0$ and $a(t_s)=0$ where $t_0$ is the time at which the collapse commences, $a_0$ being the initial value of the scalar factor and $t_s$ is the time at which the singularity forms.
We then get the time behavior of the scale factor and collapse rate as
\bea
a(t)=a_0\left[\f{t_s-t}{t_s-t_0}\right]^{n},\label{Timea}~~~~~H(t)=\f{n}{(t-t_s)},
\eea
where $t_0\leq t\leq t_s$ is the time interval of the collapse process so that, it begins at $t=t_0$, proceeds for a while and ends at $t=t_s$. In order that the scale factor vanishes at a finite amount of time we require that $n>0$. For the collapse scenario to be physically reasonable, the weak energy condition (WEC) must be respected. This condition states that the energy density as measured by any local observer is non-negative. Hence, for energy momentum tensor of ordinary matter and that of effective fluid the following conditions must be satisfied along any non-spacelike vector field
\be\label{WECrhop} \rho\geq0,~~~~\rho+p\geq0,~~~\rho_{\rm eff}\geq0,~~~~\rho_{\rm eff}+p_{\rm eff}\geq0.
\ee
Using solution (\ref{Timea}) in Eqs. (\ref{RasFES}) and (\ref{RasFES1}) we arrive at the following expressions for energy density, pressure and the second part of the above inequalities as
\bea
&&\rho(t)=\rho_0(t-t_s)^{-2},~~~p(t)=w\rho(t)\label{energy1}\\
&&\rho(t)+p(t)=(1+w)\rho(t),\label{wec3}
\eea
where
\be\label{rho0coeff}
\rho_0=\f{4(6\gamma-1)\left[3\gamma(1+w)-1\right]}{3(1+w)^2(1-4\gamma)^2}.
\ee
We also have the following expressions for effective energy density and pressure profile as
\bea
\rho_{\rm eff}(t)&=&\f{3\gamma(1+w)-1}{4\gamma-1}\rho(t),\label{rhoeffexpress}\\
p_{\rm eff}(t)&=&\f{\gamma+w(\gamma-1)}{4\gamma-1}\rho(t),\label{peffexpress}\\
\rho_{\rm eff}(t)+p_{\rm eff}(t)&=&\rho(t)+p(t).\label{rhopreffexpress}
\eea
Thus, for the sake of physical validity of the collapse setting we require that conditions (\ref{WECrhop}) hold throughout the dynamical evolution of the collapse i.e., within the time interval $t_0\leq t\leq t_s$. This gives
\bea\label{ineqsweakencon}
\rho_0\geq0,~~~~~~w\geq-1,~~~~~\f{3\gamma(1+w)-1}{4\gamma-1}\geq0.
\eea
We note that the energy density grows and diverges at $t=t_s$. This behavior along with the divergence of Kretschmann scalar
\bea\label{Kretscmann}
{\mathcal K}&=&{\mathcal R}_{\alpha\beta\delta\epsilon}{\mathcal R}^{\alpha\beta\delta\epsilon}=\dot{H}^2+2H^4+2H^2\dot{H}\nn
&=&\f{4\ell(1-3\gamma(+w))^2}{81(1+w)^4(1-4\gamma)^4(t-t_s)^4},
\eea
where
\be\label{ellco}
\ell=\left(5+6w+9w^2-36\gamma(1+w)^2+72\gamma^2(1+w)^2\right),
\ee
verifies the formation of a spacetime singularity. Next task is to examine whether the singularity that forms is hidden behind a horizon or can be detected by external observers. Basically, it is the structure of trapped surfaces during the dynamical evolution of the collapse that determines the visibility or otherwise of the spacetime singularity. These surfaces are defined as compact two-dimensional space-like surfaces such that both families of ingoing and outgoing null geodesics normal to them necessarily converge~\cite{frolovnovikov}. The interior metric (\ref{FLRW}) can be split into the surface of a 2-sphere and a two dimensional hyper-surface normal to the 2-sphere as~\cite{MSEE}
\begin{equation}\label{hmunu}
ds^2=h_{\mu\nu}dx^{\mu}dx^{\nu}+R(t,r)^2d\Omega^2,~~~h_{\mu \nu}={\rm diag}\left[-1,a(t)^2\right].
\end{equation}
Introducing the null coordinates
\begin{equation}\label{doublenull}
d\xi^{+}=-\frac{1}{\sqrt{2}}\left[dt-a(t)dr\right],~~~~d\xi^{-}=-\frac{1}{\sqrt{2}}\left[dt+a(t)dr\right],
\end{equation}
the line element (\ref{hmunu}) can be recast into double-null form as
\begin{equation}\label{metricdnull}
ds^2=-2d\xi^{+}d\xi^{-}+R(t,r)^2d\Omega^2.
\end{equation}
The radial null geodesics are given by the condition $ds^2=0$, whence, we find out that there exist two kinds of
future-directed null geodesics which correspond to $\xi^{+}=constant$ and $\xi^{-}=constant$. The expansion parameters along this geodesics are given by
\begin{equation}\label{expansion}
\Theta_{\pm}=\frac{2}{R(t,r)}\frac{\partial}{\partial\xi^{\pm}}R(t,r),
\end{equation}
where
\bea\label{p+p_}
\!\!\!\!\!\frac{\partial}{\partial\xi^{+}}\!=\!\frac{1}{\sqrt{2}}\left[\partial_{t}+\frac{1}{a(t)}\partial_{r}\right],~\frac{\partial}{\partial\xi^{-}}\!=\!\frac{1}{\sqrt{2}}\left[\partial_{t}-\frac{1}{a(t)}\partial_{r}\right].
\eea
The expansion parameter measures whether the congruence of null geodesics normal to a sphere is
diverging $(\Theta_{\pm}>0)$ or converging $(\Theta_{\pm}<0)$, that is to say, the area
radius along the light rays is increasing or decreasing, respectively. The spacetime is referred to
as trapped, untrapped and marginally trapped if
\begin{equation}\label{sp}
\Theta_{+}\Theta_{-}>0,~~~~ \Theta_{+}\Theta_{-}<0,~~~~\Theta_{+}\Theta_{-}=0,
\end{equation}
respectively, where the third class implies the outermost boundary of the trapped region, i.e., the apparent horizon.
Thus, if the apparent horizon forms early enough before the singularity formation, the singularity is covered and in case the apparent horizon fails to form or is delayed the singularity could be be visible to outside observers~\cite{Joshibook}. The key factor that determines the apparent horizon dynamics is the Misner-Sharp energy~\cite{MSE} which describes the mass contained within the shell labeled by $r$ at the time $t$ and is defined as~\cite{MSEE}
\bea\label{MS}
M(t,r)&=&\frac{R(t,r)}{2}\left[1-h^{\mu\nu}\partial_{\mu}R(t,r)\,\partial_{\nu}R(t,r)\right]\nn&=&\f{R(t,r)\dot{R}(t,r)^2}{2}.
%&=&\frac{R(t,r)}{2}\left[1+\frac{R(t,r)^2}{2}\theta_{+}\theta_{-}\right]
\eea
%\bea\label{MSEQ}
%M(t,r)&=&\f{R(t,r)}{2}\left[1-g^{\mu\nu}\partial_\mu R(t,r)\partial_\nu R(t,r)\right]\nn
%&=&\f{R(t,r)\dot{R}(t,r)^2}{2}.
%\eea
From (\ref{MS}) we can rewrite the field equations (\ref{FESEquiv})-(\ref{EFFEMT3}) in terms of an isotropic effective fluid source as\footnote{In other words, multiplying the ${ t-t}$~component of Eq. (\ref{FESEquiv}) by $R^2R^\prime$ and the ${r-r}$~component by $R^2\dot{R}$ and after a few simplification we get $\partial_r(2M)=\kappa\rho_{{\rm eff}}R^2R^\prime$ and $\partial_t(2M)=-\kappa p_{{\rm eff}}R^2\dot{R}$.}~\cite{FESEFFS}
\bea
\f{2M^\prime(t,r)}{R^2R^\prime}&=&3H^2=\kappa\rho_{\rm eff},\label{fes2meff1}\\
-\f{2\dot{M}(t,r)}{R^2\dot{R}}&=&-3H^2-2\dot{H}=\kappa p_{\rm eff},\label{fes2meff2}
\eea
where $\dot{}\equiv\partial_t$ and $\prime\equiv\partial_r$. From Eq. (\ref{fes2meff1}) we have
\be\label{2Mrhoeff}
2M(t,r)=\f{\kappa}{3}\rho_{\rm eff}R(t,r)^3.
\ee
Using equation (\ref{expansion}) one can easily check that $h^{\mu\nu}\partial_{\mu}R(t,r)\,\partial_{\nu}R(t,r)=-R^2(t,r)\Theta_{+}\Theta_{-}/2$, from which one has
\be\label{thetamissh}
\Theta_{+}\Theta_{-}=\frac{2}{R^2(t,r)}\left[\frac{2M(t,r)}{R(t,r)}-1\right].
\ee
Therefore, conditions (\ref{sp}) can be re-expressed in terms of the ratio $2M(t,r)/R(t,r)$ as 
\be\label{condstrappedness}
\f{2M(t,r)}{R(t,r)}>1,~~~\f{2M(t,r)}{R(t,r)}<1,~~~\f{2M(t,r)}{R(t,r)}=1,
\ee
where the equality provides us with the location of apparent horizon. %thus the apparent horizon
%curve is simply given by the condition $\dot{R}(t,r)^2=1$. In other words, the vector $\partial^{\mu}R$
%is space-like (time-like) in untrapped (trapped) regions and null on the apparent horizon. The spacetime is then said to be trapped, untrapped and marginally trapped if the following conditions hold, respectively
Using then expressions (\ref{EFFEMT1}), (\ref{Timea}) and (\ref{energy1}) along with (\ref{2Mrhoeff}) we get
\be\label{2MoverR}
\f{2M(t,r)}{R(t,r)}=\f{4r^2a_0^2(6\gamma-1)\left[1-3\gamma(1+w)\right]^2}{9(1+w)^2(4\gamma-1)^3(t_0-t_s)^2}\left[\f{t-t_s}{t_0-t_s}\right]^{m-1},
\ee
where
\be\label{2MoverRt}
m=\f{3w-1}{3(1+w)(4\gamma-1)}.
\ee
Now, if at initial epoch we have the ratio $2M(t_0,r)/R(t_0,r)<1$ as required by regularity of the initial conditions~\cite{Joshibook}, then for $m>1$, this ratio will remain less than unity till the singularity formation and thus the apparent horizon will not form. For $m<1$, as the collapse proceeds we reach the time $t_0<t_{\rm ah}<t_s$ so that $2M(t_{\rm ah},r)/R(t_{\rm ah},t)>1$ and thus the apparent horizon forms to cover the singularity. Figure (\ref{FIGWG1}) presents the allowed regions for a physically reasonable collapse setting in the parameter space constructed by $(\gamma,w)$. The shaded region corresponds to those values of EoS and Rastall parameters for which, a homogeneous perfect fluid collapse, $i)$ leads to the formation of a spacetime singularity (i.e., $n>0$), $ii)$ the WEC holds throughout the collapse process, i.e., the inequalities given in (\ref{ineqsweakencon}) hold, $iii)$ the regularity of initial conditions is respected, i.e., initial matter profiles do not present any singularities and are well behaved, $iv)$ the ratio $2M/R$ be less than unity at the onset of the collapse and $v)$ there is no apparent horizon to dress the singularity. Thus the collapse setting with any pair of $(\gamma,w)$ picked up from the shaded region could possibly results in the formation of a naked singularity. Moreover, if we choose the EoS and Rastall parameters from the gray region, an apparent horizon will form to cloak the singularity and thus the collapse process for this case leads to black hole formation. For $\gamma=0$, the formation or otherwise of the apparent horizon will be decided by those values of EoS parameter for which $w>-1/3$ or $w<-1/3$ respectively. We note that these conditions on EoS parameter hold only on the blue dashed line where we actually deal with GR. We further note that these conditions does not hold for $\gamma\neq0$, e.g., for $0<\gamma<0.15$, formation or otherwise of the apparent horizon depends on the value of EoS parameter and this situation also occurs for negative values of $\gamma$.
%\rc{An important issue in the collapse scenario is to consider the whole spacetime of the collapsing body, since in standard stellar collapse models the surface of the object divides the whole spacetime into two different regions: the interior region filled with matter and radiation and the exterior one which can be a region of without matter (but possibly radiation). Therefore, in order to complete the collapse model these two regions must be matched smoothly across the surface of the collapsing object.}
\section{Exterior solution}\label{exteriorsol}
As in the standard stellar collapse models the surface of the object divides the whole spacetime into two different regions, i.e., the interior and exterior ones, it is important to investigate how these regions are connected to each other. This allows us to have a complete description of the collapse setting and to this aim, the interior and exterior spacetimes must be matched smoothly across the surface of the collapsing body. A suitable candidate for exterior region is the Vaidya spacetime which has been extensively studied in gravitational collapse process~\cite{Joshibook} and particularly to describe the formation of naked singularities~\cite{nakedvaidya}. Let us take the line element for exterior spacetime as that of a generalized Vaidya metric given as~\cite{GenVaidya}
\be\label{extmetric}
ds_{+}^2=-f(\tilde{R},v)dv^2-2dvd\tilde{R}+\tilde{R}^2d\Omega^2,
\ee
where $f(\tilde{R},v)=1-2{\mathcal M}(\tilde{R},v)/\tilde{R}$ is the exterior metric function with ${\mathcal M}(\tilde{R},v)$ being a measure of mass enclosed within the radius $\tilde{R}$ and $v$ being the retarded null coordinate. Using Israel-Darmois junction conditions~\cite{ISRAEL}, the above metric can be matched to the interior line element (\ref{FLRW}) through the boundary surface given by $r=r_b$. The interior and exterior reduced metrics then take the form, respectively~\cite{Poissonbook}
\bea
ds^2_{-}|_{r=r_b}&=&-dt^2+a(t)^2r_b^2(d\theta^2+\sin^2\theta d\phi^2),\label{intreduced}\\
ds^2_{+}|_{r=r_b}&=&-\left[f\left(\tilde{R}(t),v(t)\right)\dot{v}^2+2\dot{\tilde{R}}\dot{v}\right]dt^2\nn
&-&\tilde{R}^2(t)\left(d\theta^2+\sin^2\theta d\phi^2\right),\label{extreduced} 
\eea
whereby matching the induced metrics give
\be\label{matchintext}
f\left(\tilde{R}(t),v(t)\right)\dot{v}^2+2\dot{\tilde{R}}\dot{v}=1,~~~\tilde{R}(t)=r_ba(t).
\ee
A straightforward but lengthy calculation reveals that the non-vanishing components of the extrinsic curvature of hypersurface $r=r_b$ for interior and exterior regions are found as~\cite{matchingdetails} 
\bea
\!\!\!\!\!\!\!\!\!\!\!\!\!\!\!\!\hspace{-1.2cm}K_{tt}^{+}\!\!=\!\!&-&\!\!\!\!\frac{\dot{v}^2\left[ff_{,\tilde{R}}\dot{v}+f_{,v}\dot{v}+3f_{,\tilde{R}}\dot{\tilde{R}}\right]+2\left(\dot{v}\ddot{\tilde{R}}-\dot{\tilde{R}}\ddot{v}\right)}{2\left(f\dot{v}^2+2\dot{\tilde{R}}\dot{v}\right)^{\frac{3}{2}}},\label{Kttext}\\
\!\!\!\!\!\!\!\!\!\!\!\!\!\!\!\!\hspace{-1.2cm}K^{{+}\theta}_{\theta}\!\!\!\!&=\!\!\!\!&K^{{+}\phi}_{\phi}\!\!=\!\!\frac{f\dot{v}+\dot{\tilde{R}}}{\tilde{R}\sqrt{f\dot{v}^2+2\dot{\tilde{R}}\dot{v}}},\nn\label{Kthhthhext}
K_{tt}^{-}&=&0,~~~~K^{-\theta}_{\theta}=K^{-\phi}_{\phi}=\frac{1}{r_{b}a(t)}\label{intextinsicc}.
\eea
Matching the extrinsic curvatures along with using (\ref{matchintext}) we find that $\partial_v f(\tilde{R},v)=0$ implying that the exterior metric function must be independant of retarded null coordinate. Furthermore, the four velocity of the boundary as seen by an exterior observer is obtained as
\be\label{4Velocity}
u^\alpha=\left(\dot{v},\dot{\tilde{R}},0,0\right)=\left[\f{1+\sqrt{1-f}}{f},-\sqrt{1-f},0,0\right],
\ee
where the minus sign stands for a collapse scenario. From equation (\ref{fes2meff1}) and the second component of the above vector field, we get for a smooth matching ${\mathcal M}(\tilde{R})=M(t,r_b)$, whence we have
\be\label{matchingmasses}
\f{2{\mathcal M}}{\tilde{R}}=\f{n^2r_b^{\f{2}{n}}}{(t_0-t_s)^2}\tilde{R}^{\f{2}{n}(n-1)}.
\ee
Therefore the line element for exterior spacetime reads
\be\label{lineelext}
ds_+^2=-\left(1-2{\mathcal M}_0\tilde{R}^{\f{2}{n}(n-1)}\right)dv^2-2dvd\tilde{R}+\tilde{R}^2d\Omega^2,
\ee
where ${\mathcal M}_0=n^2r_b^{\f{2}{n}}/2(t_0-t_s)^2$. We note that the exterior solution for $\gamma=0$ and $w=0$ is a Schwarzschild metric with dynamical boundary in retarded null coordinates.
\section{concluding remarks}
In the present work, the collapse process of a homogeneous perfect fluid was studied in Rastall gravity and it was shown that, depending on the values of EoS and Rastall parameters, collapsing configurations from regular initial data could end in either a black hole or naked singularity formation. Whereas the Rastall parameter can be regarded as the strength of mutual non-minimal coupling between matter and geometry, we observed that such a coupling could affect the formation of apparent horizon so that it is possible to have naked singularities in homogeneous perfect fluid collapse for $w<-1/3$ instead of black hole formation which occurs for these values of EoS parameter in GR. Hence, these solutions in Rastall gravity may be considered as counter-examples to CCC. Moreover from the effective fluid viewpoint, if we consider an effective dust collapse, i.e., $p_{\rm eff}=0$, we then expect black hole formation as the collapse end product. However, in this case, ordinary matter profiles have to obey the EoS $p=w\rho=\gamma/(1-\gamma)\rho$, i.e., the pair $(\gamma,w)$ have to lie on the solid curve in Fig. (\ref{FIGWG1}). We then conclude that, even if the effective pressure vanishes, black hole formation could occur for non-vanishing fluid pressure, the situation which is absent within the framework of GR for gravitational collapse of a homogeneous dust cloud. 
\acknowledgments{The authors would like to thank the anonymous referee for useful comments and suggestions that helped us to improve our manuscript. The work of A. H. Ziaie has been supported financially by Research Institute for Astronomy \& Astrophysics of Maragha (RIAAM).}

\end{document}